\documentclass[fleqn,twoside]{article}
\usepackage{espcrc2,epsf}

\title{$I=2$ $\pi\pi$ scattering using G-parity boundary condition
\thanks{This work was supported by in part by
the U.~S.~Department of Energy and the RIKEN BNL Research Center.}}

\author{Changhoan Kim, \address{Department of Physics, Columbia
University, New York, NY, 10027}}
\begin{document}

\bibliographystyle{apsrev}

\begin{abstract}
To make the $\pi\pi$ state with non-zero relative momentum as the leading exponential,
we impose anti-periodic boundary condition on the pion, which
is implemented by imposing G-parity or H-parity 
on the quark fields at the boundary.
With this, we calculate the $I=2$ $\pi\pi$ phase shift 
from lattice simulation by using L\"uscher's formula.

\end{abstract}

\maketitle

\section{Introduction}
\label{sec:Intro}
Lattice gauge theory provides a way to investigate low energy physics
of QCD, which cannot be done using any perturbative method.  
One of the interesting physical quantities is 
related to $\pi\pi$ system. 
The $K \rightarrow \pi\pi$ WME which violates $CP$ symmetry
is of particular interest.
Since lattice calculations easily extract ground state,
the need to generate the final $\pi\pi$ state with 
non-trivial relative momentum is a serious difficulty. 
We proposed to use G-parity boundary conditions to overcome
this difficulty{\cite{Kim:2002np}.
In the present paper, we have implemented this G-parity idea and a new H-parity boundary condition
in numerical simulations and 
calculated the $I=2$, $\pi\pi$ phase shift.


\section{G parity boundary condition}

Since the G-parity operation on a pion gives 
\begin{equation}
G | \pi^\pm > = - | \pi^\pm > ~,~ G | \pi^0 > = - | \pi^0 >, \nonumber
\end{equation}
by applying this operation on the boundary, we can impose 
anti-periodic boundary condition on pion. 
To implement this condition on lattice, 
we have to use the 
G-parity operation on the quark fields:
\begin{equation}
G\left( \begin{array}{c} {\mathbf{u}} \\[2mm] {\mathbf{d}} \end{array} \right)
= 
\left(
\begin{array}{c} -{\mathbf{d}}^C \\[2mm] {\mathbf{u}}^C \end{array}
\right). \nonumber
\end{equation} 
In actual calculation, we impose this boundary condition 
only in the $z$-direction so that we have a pion with non-zero $z$-momentum.
Because at the boundary there are terms such as $\psi\psi$ and $\bar{\psi} \bar{\psi}$,
it requires some special care to implement~\cite{Wiese:1992ku}. 
First, we have to impose a charge-conjugate boundary condition on the gauge field
to keep gauge invariance and we have to virtually double the box size for the
Dirac operator inversion.
Since isospin has an important role in the two pion system, it is worth noting that
G-parity commutes with isospin, which means that under this unusual boundary
condition isospin is still a good quantum number.

\section{H parity boundary condition}

An easier way to impose anti-periodic boundary conditions on a pion is
to apply the following operation on the quark fields,
\begin{equation}
H\left( \begin{array}{c} {\mathbf{u}} \\[2mm] {\mathbf{d}} \end{array} \right)
= 
\left(
\begin{array}{c} -{\mathbf{u}} \\[2mm] {\mathbf{d}} \end{array} \right). \nonumber
\end{equation}
We will call this operation H-parity .
Then, the operation on the pions will be,
\begin{equation}
H | \pi^\pm > = - | \pi^\pm > ~,~ H | \pi^0 > = | \pi^0 >. \nonumber
\end{equation}
Under this boundary condition, isospin is
not a good quantum number anymore, but $I_z$ is still good. 
Since we know that the $ I_z=2, \pi\pi $ state is composed of two $\pi^+$, 
this state must have non-zero relative momentum.
This is not true for the $I=0$ state.
So the utility of this boundary condition is more limited than the G-parity
boundary condition. However it has the advantage that it doesn't require any modification of
the gauge field boundary condition.
This allows us to use existing lattices, including dynamical ones. 

\section{Results}

\subsection{Single pion}
We first investigate the properties of the one-pion system.
We expect to find a one-pion state with momentum $\pi \over L$,
unlike the conventional $2\pi \over L$.
Figure \ref{single_pion_plot} shows a graph of energy versus pion mass.
As expected, the single pion state with smallest energy
has $E(m_\pi)=\sqrt{m_\pi^2 + sin^2(\frac{\pi}{L}})$.

\begin{figure}
\epsfxsize=\hsize
\begin{center}
\epsfbox{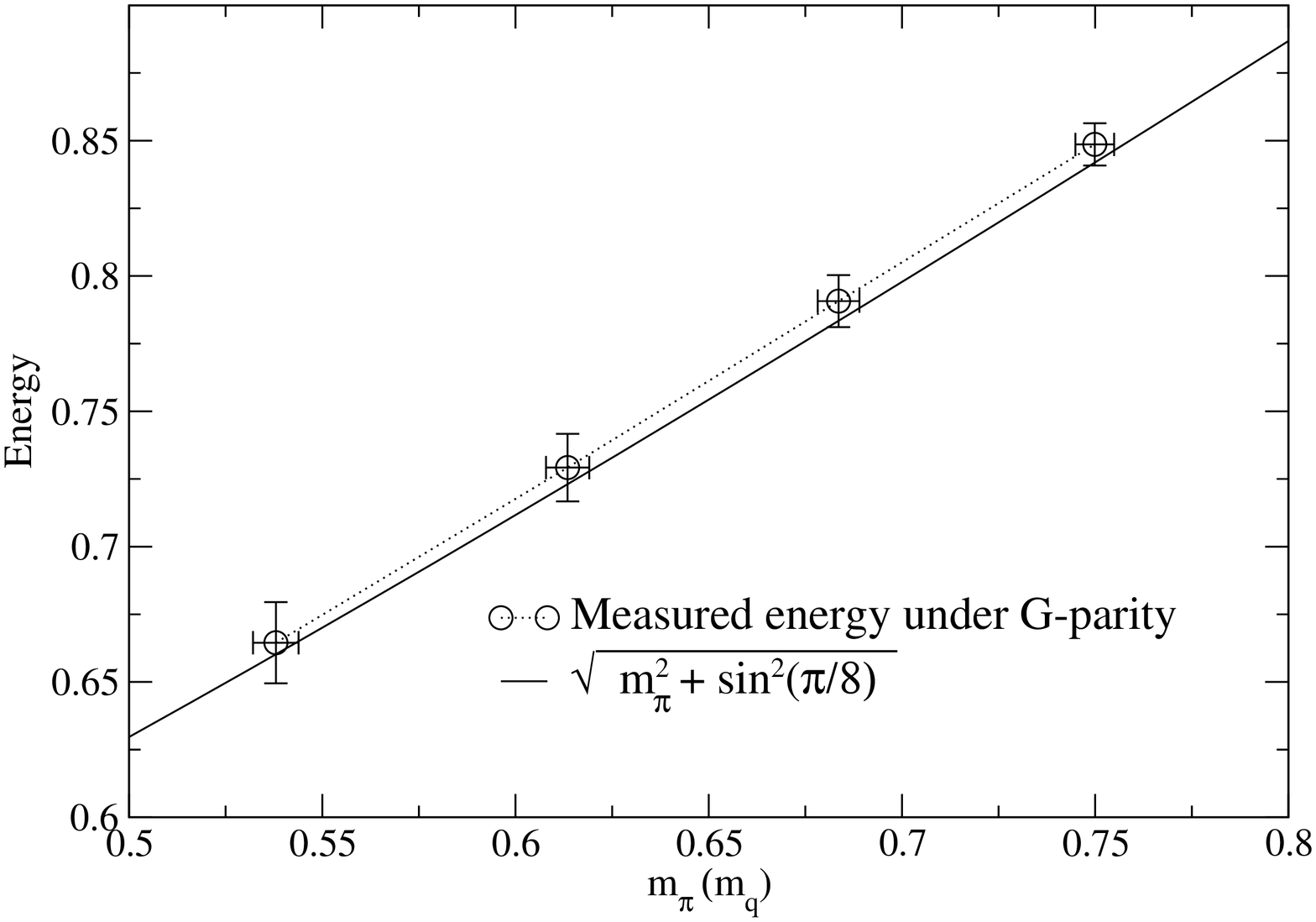} 
\vspace{-0.6in}
\caption{$1/a =$ 0.978(14)GeV, $\beta=$5.7, $8^3 \times 32$, $L_s$=10,  Wilson Gauge Action, Domain Wall Fermions (DWF),
H-parity boundary condition.\label{single_pion_plot}}
\end{center}
\vspace{-0.4in}
\end{figure}

\subsection{Two pions}

\begin{figure}
\epsfxsize=\hsize
\begin{center}
\epsfbox{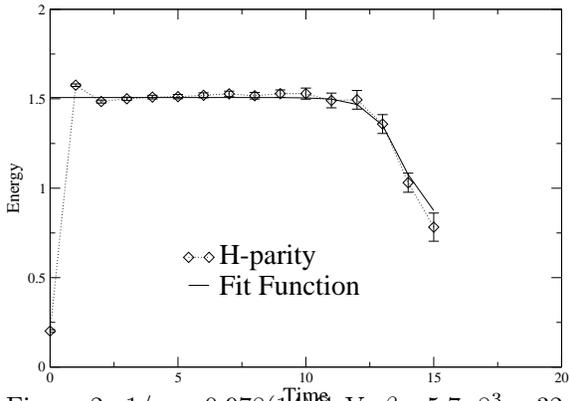} 
\vspace{-0.5in}
\caption{$1/a =$ 0.978(14)GeV, $\beta=$5.7, $8^3 \times 32 $, $L_s$=10,  Wilson Gauge Action, DWF, 
H-parity.} \label{H_parity_eff_mass2}
\end{center}
\vspace{-0.4in}
\end{figure} 

Figure \ref{H_parity_eff_mass2} shows the effective mass plot for the two pion state.
We have a very nice plateau in the time range from 3 to 11,
but after that we have suspicious fall-off.
This fall-off can be explained by considering Fig.~\ref{pion_prop}.
A pion created at $t=0$ can propagate in either direction.
Because we have two particles, there is a state in which the two particles propagate
in opposite directions. 
The correlation function for this state is
\begin{equation}
G(t) \approx e^{-E_\pi(T-t)} e^{-E_\pi t } = e^{-E_\pi T }=  e^{-2 E_\pi \frac{T}{2} }  
\end{equation}
here $E_\pi$ means the energy of one pion with momentum.
This is just a constant, and will be dominant near $t=\frac{T}{2}$
because the energy of the $I=2$, two-pion state is bigger than $2 E_\pi$
and can cause the abrupt fall-off in this region.
We confirmed this idea by fitting the effective mass plot with the function
``$cosh + const.$'', the solid line in Fig.~\ref{H_parity_eff_mass2}.

\begin{figure}
\epsfxsize=\hsize
\begin{center}
\epsfbox{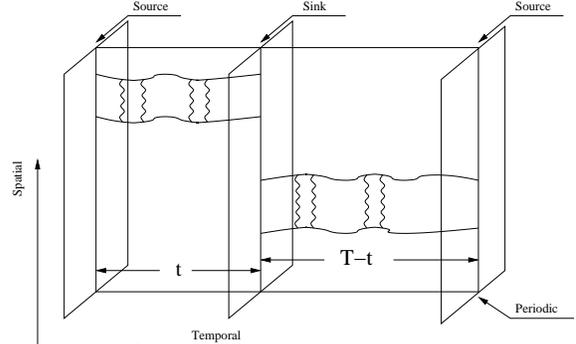}
\vspace{-0.5in}
\caption{One pion state exists all time.} \label{pion_prop}
\end{center}
\vspace{-0.4in}
\end{figure}

Figure \ref{H_parity_vs_G_parity} shows the same two-pion state effective mass plot for the G-parity boundary condition.
This G-parity effective mass plot is quite different from the one for H-parity.
Instead of an abrupt fall off, it has gradual decrease.
It even looks like it has two plateaus.
A simulation with more time slices ($N_t=48$) also shown in Fig.~\ref{H_parity_vs_G_parity}  demonstrates this two-plateau structure.
Since the spatial lattice volume was small ($\approx (1.7fm)^3$) for this calculation,
we guessed that it might be a finite volume effect and performed the same simulation with a bigger volume.
Figure \ref{8816} shows the result for a spatial volume $\approx 1.7fm \times 1.7fm \times 3.4fm$.

\begin{figure}[t]
\epsfxsize=\hsize
\begin{center}
\epsfbox{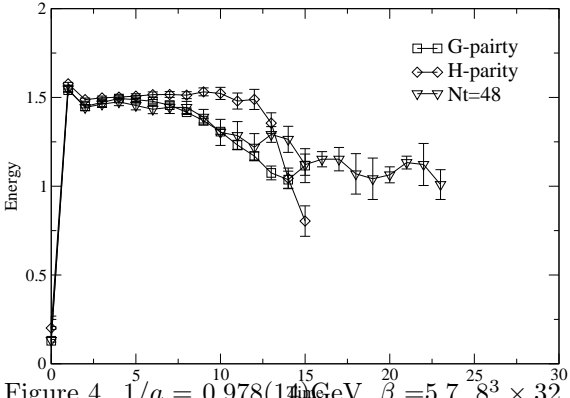}
\vspace{-0.55in}
\caption{$1/a =$ 0.978(14)GeV, $\beta=$5.7, $8^3 \times 32 $, $L_s$=10, Wilson Gauge Action, DWF} \label{H_parity_vs_G_parity}
\end{center}
\vspace{-0.4in}
\end{figure}

\begin{figure}
\epsfxsize=\hsize
\begin{center}
\epsfbox{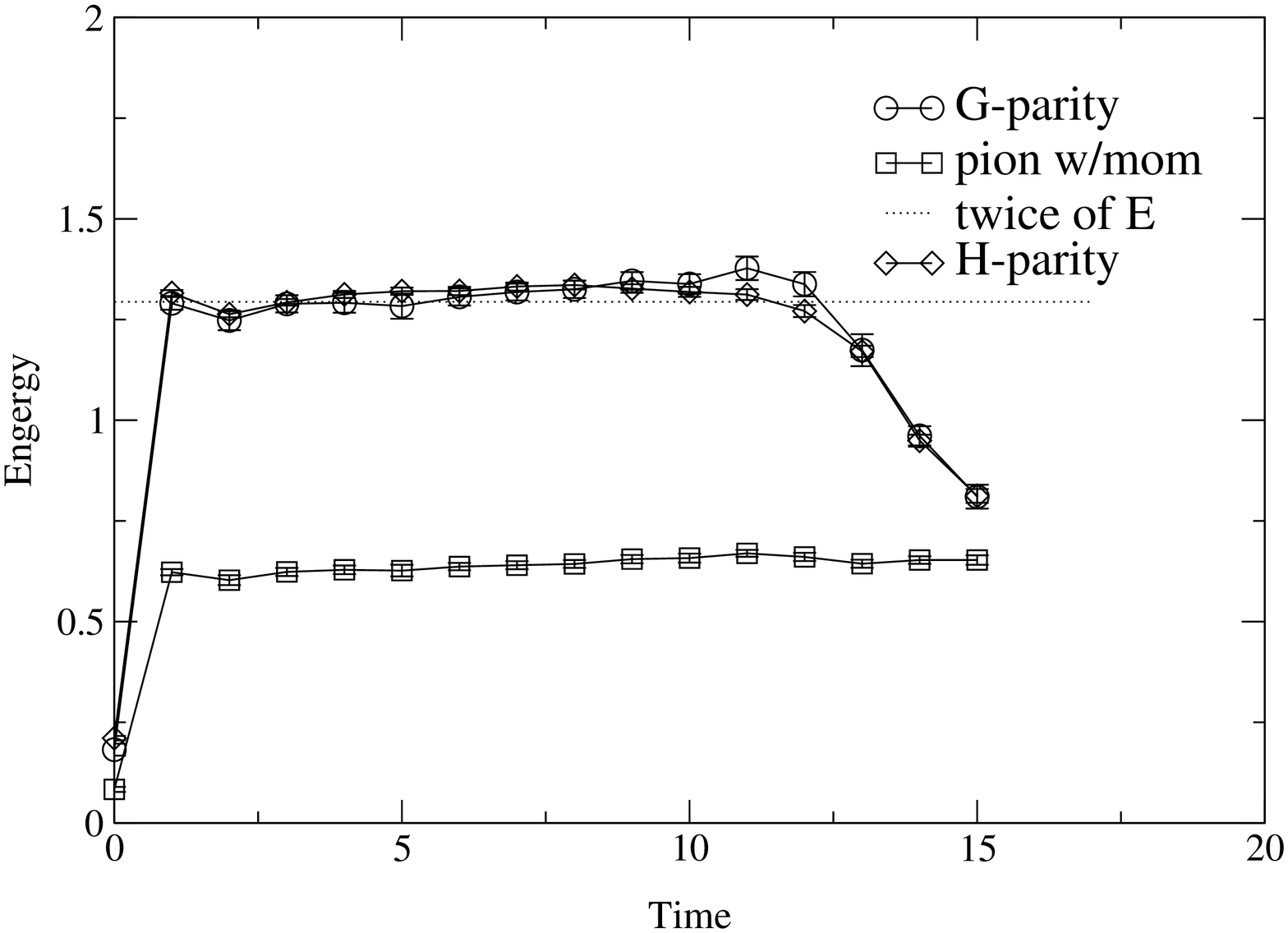} 
\vspace{-0.5in}
\caption{$1/a =$ 0.978(14)Gev, $\beta=$5.7, $8\times 8\times 16\times 32$,
 $L_s$=10, Wilson Gauge Action, DWF} \label{8816}
\end{center}
\vspace{-0.4in}
\end{figure}

We notice that the gradual decrease of G-parity plot has disappeared and
it is almost identical to that seen with H-parity.
Therefore, we can conclude that it is a finite volume effect that caused
the two plateau behaviour.
This might mean that we have discovered a new finite volume $\bar{q}q \bar{q}q$ state. 
After becoming convinced from these tests that we have a two-pion state with non-zero relative 
momentum, we extended this simulation and calculated the $I=2$ $\pi\pi$ phase shift. 
Since we are using L\"uscher's formalism\cite{Luscher:1991ux}., this is nothing but spectroscopy.
Figure \ref{phase_plot} shows our phase shift calculation including CP-PACS\cite{Aoki:2002ny} and
experimental results.
The following are our $\delta_{\pi\pi}$ simulation parameters ($1/a$ is in GeV 
and $N_t$=32):
\begin{tabbing}
Domain Wall Fermions, $L_s$=10, $M_5$=1.65 \\ 
p \=$\approx$\hspace{0.1cm}250MeV \\
\>G\hspace{0.3cm}\=$8^2\times16$\hspace{0.5cm}\=$1/a=$0.978(14)\hspace{0.3cm}\=Wilson 91conf\\
\>H\>$8^2\times16$\>$1/a=$0.978(14)\>Wilson 172conf\\
p $\approx$ 450MeV \\
\>H\>$8^3$\>$1/a=$0.978(14)\>Wilson 269conf\\
\>H\>$16^3$\>$1/a=$1.98(3)\>DBW2 156conf\\
\end{tabbing}
\vspace{-0.5cm}

\begin{figure}
\epsfxsize=\hsize
\begin{center}
\epsfbox{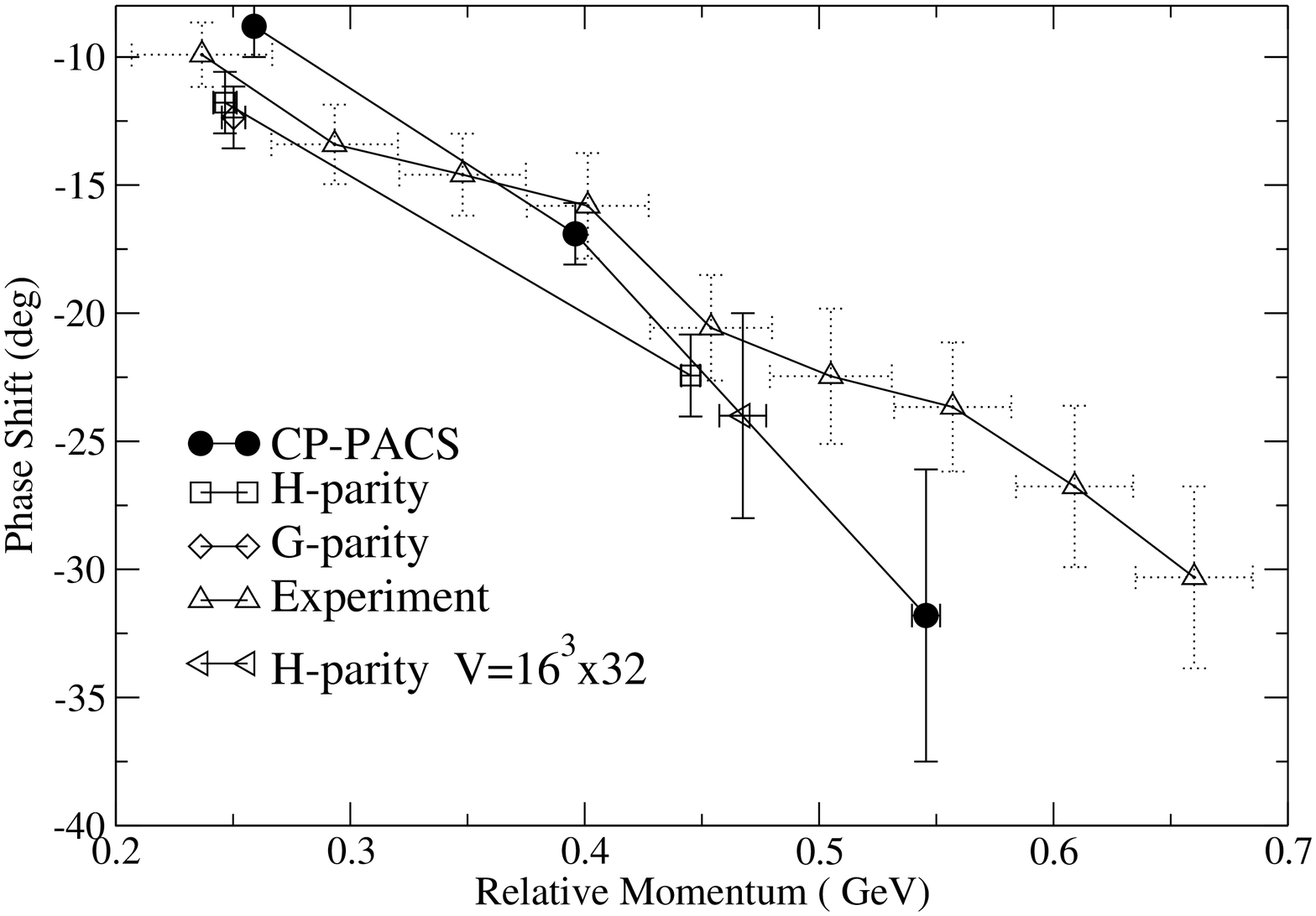} 
\vspace{-0.5in}
\caption{Phase shift results versus momentum.} \label{phase_plot}
\end{center}
\vspace{-0.4in}
\end{figure}

\section{Conclusion}
\label{sec:Conclusion}
We have tested the idea of imposing anti-periodic boundary conditions
on the pion by applying a G-parity or H-parity operation on the quark field
at the boundary.
For the one pion case, we found that the energy of the particle is given
by $\sqrt{m_\pi^2 + sin^2(\frac{\pi}{L})}$ as expected.
We can achieve a two-pion state with non-zero momentum, and
from this relative momentum, we can calculate the $I=2$, $\pi\pi$ phase shift.
Since the H-parity boundary condition can be applied to existing lattices,
it will be more convenient than G-parity.  In particular, we plan to use the H-parity boundary condition 
for a $\Delta I=\frac{3}{2}~~~ K\rightarrow\pi\pi$ calculation on existing lattices.  
However, for the $I=0$, $\pi\pi$ state only the G-parity boundary condition 
with dynamical lattices will work.  
Since the G-parity boundary condition is more vulnerable to finite
volume effects, {\it e.g.} the $\bar{q}q\bar{q}q$ state which we need to study further,
it may require more resources.

\end{document}